\documentclass[%
 reprint,
 amsmath,amssymb,
 aps,prl
]{revtex4-2}

\usepackage{graphicx}
\usepackage{dcolumn}
\usepackage{bm}
\usepackage{color}

\usepackage[normalem]{ulem}

\begin{document}

\preprint{APS/123-QED}

\title{Coupling Neutrino Oscillations and Simulations of Core-Collapse Supernovae}

\author{Charles J.\ Stapleford}
\email{cjstaple@ncsu.edu}
\author{Carla\ Fr{\"o}hlich}
\email{cfrohli@ncsu.edu}
\author{James\ P.\ Kneller}
\email{jpknelle@ncsu.edu}
\affiliation{Department of Physics, North Carolina State University, Raleigh, NC 27695, USA}

\date{\today}

\begin{abstract}
At the present time even the most sophisticated, multi-dimensional simulations of core-collapse supernovae do not (self-consistently) include neutrino flavor transformation. This physics is missing despite the importance of neutrinos in the core-collapse explosion paradigm. Because of this dependence, any flavor transformation that occurs in the region between the proto-neutron star and the shock could result in major effects upon the dynamics of the explosion. We present the first hydrodynamic core-collapse supernova simulation which simultaneously includes flavor transformation of the free-streaming neutrinos in the neutrino transport. These oscillation calculations are dynamically updated and evolve self-consistently alongside the hydrodynamics. Using a $M=20\;{\rm M_{\odot}}$ progenitor, we find that while the oscillations have an effect on the neutrino emission and the heating rates, flavor transformation alone does not lead to a successful explosion of this progenitor in spherical symmetry.
\end{abstract}

\maketitle

{\bf Introduction}
Since the earliest simulations of core-collapse supernovae by Colgate and White \cite{1966ApJ...143..626C}, neutrinos have been recognised as important participants in the dynamics of the explosion. Accurate treatment of the neutrinos in a supernova simulation is a difficult problem because of the huge changes in their coupling to the rest of the matter content. The most sophisticated simulations solve the classical Boltzmann equation for the neutrino distribution function \cite{1966AnPhy..37..487L,2017ApJ...847..133R} but given its significant computational burden, many approximate neutrino transport schemes have been developed over the years. We refer the reader to O'Connor \emph{et al.} \cite{2018JPhG...45j4001O} and Pan \emph{et al} \cite{2019JPhG...46a4001P} for a comparison of various 1D and 3D simulation codes which use different neutrino transport schemes.  

The fundamental assumption underlying all classical neutrino transport schemes is that each neutrino flavor is conserved separately. This is clearly incorrect and there is a wealth of evidence from experiments that neutrino flavors mix. In order to include this fundamental physics in the simulations, one must treat the neutrino as a quantum particle. This necessity has been recognized for some time but little progress has been made in its direction. Firstly, a quantum treatment of neutrinos is computationally far more expensive than a classical treatment. Secondly, post-processed flavor transformation calculations of supernova simulations (which had used classical treatments) indicated the flavor transformation should be suppressed during the accretion phase and should not lead to significant changes to the dynamics \cite{2011PhRvL.107o1101C,2011ApJ...738..165S,2012PhRvD..85f5008D}. For self-consistency, however, one must consider the feedback of flavor oscillations upon the supernova dynamics.

In addition to self-consistency, there are two additional strong motivations to go beyond post-processing calculations and include flavor transformation in simulations. First, recent developments in the understanding of flavor oscillations in supernovae gives one reason to believe flavor oscillations may start much closer to the proto-neutron star than indicated in those earlier, post-processed studies. So-called `fast oscillations' \cite{2005PhRvD..72d5003S,2012PhRvL.108w1102M,2012PhRvD..85k3002S,2016PhRvL.116h1101S,2017JCAP...02..019D,2017arXiv171207013A,2017PhRvL.118b1101I,Azari2019FastSupernovae,Johns2019NeutrinoInstabilitiesb,Yi2019TheWave} may occur as a result of `angular crossings' in the electron neutrino/antineutrino spectra --- the electron lepton number current changes sign as a function of the angle relative to the radial direction --- at radii below the shock. Second, Beyond the Standard Model physics can also lead to significant changes to the flavor evolution close to the proto-neutron star \cite{1999NuPhB.538..368B,1999PhRvC..59.2873M,2006PhRvD..73i3007B,2008PhRvD..78k3004B,2012JCAP...01..013T,2014PhRvD..89f1303W,2014PhRvD..90c3013E,2016PhRvD..94i3007S}. Whatever the origin of the flavor oscillations, if they occur well below the shock the effect upon the dynamics would be significant and, in addition, have an impact upon the electron fraction which will affect the resulting nucleosynthesis.

In this paper we present the first core-collapse simulation which computes the flavor transformation simultaneously with the hydrodynamics in order to determine the feedback of flavor mixing upon the explosion and the neutrino emission. After describing how the coupling was achieved, we then determine the effect of the oscillations upon the dynamics of the explosion using a 20 $M_{\odot}$ progenitor. We finish by presenting our findings, the limitations of our study, and the directions for further improvement.\newline


\begin{figure}
    \centering
    \includegraphics[width=0.975\linewidth]{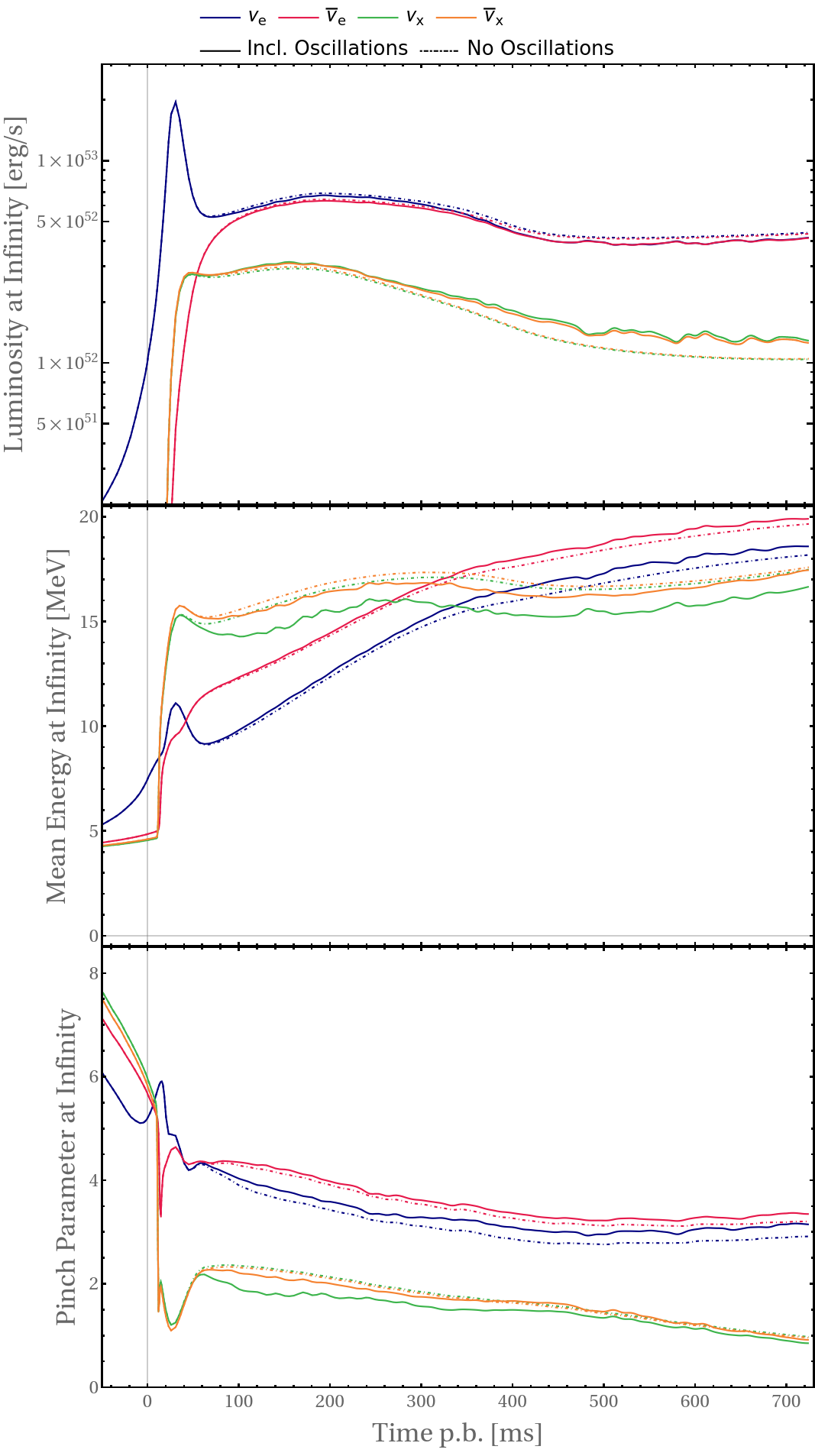}\\
    \caption{\label{fig:Nuplot} The luminosity (top), mean energy (center), and pinch parameter (bottom) for all four neutrino flavors as calculated at infinity. The results including oscillations (solid) are compared with the simulation with no oscillations (dot-dashed).
    }
\end{figure}


{\bf Including Oscillations} The feedback of flavor oscillations upon the explosion dynamics was achieved by the synthesis of the neutrino oscillation code \texttt{SQA}, and the hydrodynamical code \texttt{Agile-BOLTZTRAN}. \texttt{Agile-BOLTZTRAN} itself is the combination of 1D Lagrangian hydrodynamics, \texttt{Agile}, with classical Boltzmann neutrino transport, \texttt{BOLTZTRAN}. The hydrodynamics is calculated in general relativistic space-time on an adaptive grid to ensure high resolution where it is needed. The transport implicitly solves the finite differenced  O(v/c) Boltzmann equation for neutrinos and antineutrinos separately with two neutrino flavors (electron-type and x-type, representing both $\mu$ and $\tau$ flavors) for each i.e. it evolves four neutrino spectra in total. 
A full description of the \texttt{Agile-BOLTZTRAN} code can be found in Liebend\"{o}rfer \emph{et al.} \cite{Liebendoerfer:2002xn}. For our simulations we use 20 energy groups, 8 angle bins, and 145 radial zones.

The complete quantum description of neutrinos including flavor oscillations and coupling to matter is known as the Quantum Kinetic Equations \cite{SIGL1993423,2013PhRvD..87k3010V,2014PhRvD..89j5004V,CIRIGLIANO201527}. At the present time these equations are too computationally expensive to be implemented in full supernova simulations. If we make the assumption that dense matter and frequent collisions strongly suppress flavor mixing, we only need to treat the flavor evolution of the free-streaming neutrinos. This is much more computationally feasible and our chosen approach. \texttt{SQA} is a multi-energy, free-streaming neutrino oscillation code for three neutrino and three antineutrino flavors which assumes steady-state conditions.
It solves the same neutrino flavor evolution problem as the two codes described in Duan, \textit{et al.}~\cite{Duan:2006an} but makes use of the theoretical developments found in Galais \textit{et al.}~\cite{Galais:2011jh}.
\texttt{SQA} solves, for each energy, the Schr\"{o}dinger Equation for the evolution operators $S(r,r_0)$ and $\bar{S}(r,r_0)$ which relate the neutrino and antineutrino wave functions respectively at some initial point $r_0$ to the wave function at $r>r_0$ using an adaptive 6th order Runge-Kutta integrator. The flavor evolution is governed by the Hamiltonian which includes contributions from the vacuum potential, the Mikheyev-Smirnov-Wolfenstein (MSW) potential due to neutrino interactions with matter \cite{Mikheyev:1985aa,Mikheyev:1986tj,1978PhRvD..17.2369W}, as well as the neutrino-neutrino interaction (self-interaction) potential \cite{1995PhLB..342..250P,Duan:2006an}. General relativistic corrections to the Hamiltonian are included \cite{Yang:2017asl}. Due to its lower computational cost, we employ the single-angle approximation. We use the Particle Data Group's best fit values for the neutrino mixing angles and mass splittings~\cite{PhysRevD.98.030001}, specifically $\Delta\rm{m}_{21}^2 = 7.59\times10^{-5}$, $\Delta\rm{m}_{32}^2 = -2.43\times10^{-3}$, $\theta_{12} = 34.4^{\circ}$, $\theta_{13} = 9^{\circ}$, $\theta_{23} = 45^{\circ}$. We consider an inverted mass hierarchy only because of the better qualitative agreement between single-angle and multi-angle calculations for this ordering: see, for example \cite{2011PhRvL.106i1101D,2018JPhG...45d3002H} for details about this issue with the single angle approximation. The \texttt{SQA} calculation uses 150 energy bins subdivided from the \texttt{Agile-BOLTZTRAN} energy groups so that we maintain 0.5 MeV resolution below $\sim$50 MeV, 1 MeV resolution up to $\sim$120 MeV, and 8 MeV resolution up to 300 MeV, the maximum energy considered for the neutrino transport.

The time evolution of our simulation is controlled by \texttt{Agile-BOLTZTRAN}. As it proceeds, \texttt{SQA} is periodically invoked. \texttt{SQA} calculates the MSW potential from the mass density and electron fraction provided by \texttt{Agile-BOLTZTRAN}, and constructs the neutrino spectra for each flavor from the neutrino luminosity, mean energy, and mean square energy, also provided by \texttt{Agile-BOLTZTRAN}, assuming a `pinched spectrum' \cite{2003ApJ...590..971K}. \texttt{SQA} then computes the evolution operators $S_{i,k}$ and $\bar{S}_{i,k}$ for neutrinos and antineutrinos across each spatial zone $i$ it was given, and each energy $k$. \texttt{SQA} will return to \texttt{Agile-BOLTZTRAN} the transition probabilities $P_{i,k}$ and $\bar{P}_{i,k}$ which are the probability for electron neutrino or electron antineutrinos to be converted. These probabilities are given by $P_{i,k} = 1-|S_{i,k;ee}|^2$ and $\bar{P}_{i,k} = 1-|\bar{S}_{i,k;ee}|^2$ where $S_{i,k;ee}$ is the `ee' element from the S matrix (and similar for $\bar{S}_{i,k;ee}$). We take advantage of the fact that the `ee' element is the same for both 3-flavor (\texttt{SQA}) and 2-flavor (\texttt{Agile-BOLTZTRAN}) oscillation calculations to define our transition probabilities. Note that we do not assume flavor diagonal density matrices at the beginning of each zone: instead the `initial' neutrino density matrix $\rho_{i+1,k}$ for the zone $i+1$ is related to the `initial' neutrino density matrix of the previous zone via $\rho_{i+1,k} = S_{i,k}\, \rho_{i,k}\, S^{\dagger}_{i,k}$. Only at the initial point $r_0$, taken to be $10\;{\rm km}$ above the neutrinosphere, are the density matrices set to be diagonal in the flavor basis. The onset of neutrino oscillations $10\;{\rm km}$ above the neutrinosphere was selected based on the neutrino mean free path as calculated by \texttt{Agile-BOLTZTRAN}.

\texttt{Agile-BOLTZTRAN} converts the transition probabilities $P_{i,k}$ and $\bar{P}_{i,k}$ into effective opacities for neutrino and antineutrinos, $\sigma_{i,k}$ and $\bar{\sigma}_{i,k}$, by multiplying by the speed of light $c$ and dividing by the width of the zone $r_i-r_{i-1}$, i.e. 
\begin{equation}
\sigma_{i,k} = \frac{P_{i,k}\, c}{r_i-r_{i-1}}, \hspace{20 pt} \bar{\sigma}_{i,k} = \frac{\bar{P}_{i,k}\,c}{r_i-r_{i-1}}.
\end{equation}
Since oscillations must conserve total neutrino number, we link `absorption' due to oscillations in one flavor to `emission' in the other flavor resulting in the following coupled differential equations for the distribution functions $f$ of neutrino flavors $\alpha$ and $\beta$:
\begin{equation}
    \frac{df_{\alpha}}{dt} = \sigma_{i,k}(f_{\beta} - f_{\alpha}), \hspace{20 pt} \frac{df_{\beta}}{dt} = \sigma_{i,k}(f_{\alpha} - f_{\beta}),
\end{equation}
and similarly for the antineutrinos.
These additional flavor changing terms are added to the neutrino transport equations for radially outward moving neutrinos only, using the same finite differencing scheme the code uses for other emission and absorption processes. All inward moving neutrinos are assumed to be unaltered by the oscillations. This assumption omits the neutrino halo effect \cite{2012PhRvL.108z1104C,2013PhRvD..87h5037C} but, because the density in the core region is high, we do not expect the halo effect to be significant due to matter suppression \cite{2011PhRvL.107o1101C,2012PhRvL.108f1101S,PhysRevD.85.113007}. The oscillation calculations are only executed in the post-bounce (p.b.) period after the shock has traveled beyond the neutrinosphere because only then will there be free-streaming neutrinos behind the shock where the oscillation calculations could affect the heating. 

We define an opacity correction term $\delta\sigma_{i,k}$ to track the effect of the changing size of the zones in the adaptive grid between calls to \texttt{SQA}:
\begin{equation}
    \delta\sigma_{i,k} = \sigma_{i,k}\,\frac{r_i-r_i^o}{r_i^o - r_{i-1}^o}. 
\end{equation}
In this equation $r_i$ is the radius of zone $i$ and $r_i^o$ is the radius of zone $i$ the last time \texttt{SQA} was invoked. The oscillation opacities are updated at least every 10 ms of simulation time, but will update earlier if any $\delta\sigma_{i,k}$ or $\delta\bar{\sigma}_{i,k}$ is larger than 10\% for any zone $i$ or energy $k$. 
\newline 


{\bf The Effects of Oscillations}
With our new \texttt{Agile-BOLTZTRAN-SQA} code we are able to self-consistently include the effects of flavor transformations of free streaming neutrinos upon the hydrodynamics and the altered hydrodynamics upon the neutrino emission. To explore this feedback we adopt the 20 $M_{\odot}$ progenitor from Woosley \& Heger \cite{2007PhR...442..269W}, the same progenitor used for the comparison of 1D core-collapse supernova simulations by O'Connor \emph{et al.} \cite{2018JPhG...45j4001O}. 

We performed two simulations, one including neutrino oscillations as described above, and one without neutrino oscillations for comparison. The neutrino transition probabilities we calculated were all $<\,1\%$ over the width of a single zone. This agrees with previous assumptions about the scale of neutrino flavor transformation inside the shock radius; however, these changes do still produce some noticeable differences in our simulation results. 

In figure~\ref{fig:Nuplot} we show the effects of neutrino oscillations on the luminosity, mean energy, and pinch parameter for both electron and x-type (anti)neutrinos. The overall trend is that the oscillations increase the average energy, and pinch parameter of the electron (anti)neutrinos as the higher-energy x-type (anti)neutrinos change flavor; however, the luminosity of electron (anti)neutrinos decreases significantly as there are more neutrinos which undergo $e\to x$ flavor transitions than undergo $x\to e$ transitions. Overall, the electron (anit)neutrino luminosities drop by $\sim3-5$\%, while their mean energies increase by $\sim2-3$\%.

\begin{figure}
    \centering
    \includegraphics[width=0.45\textwidth]{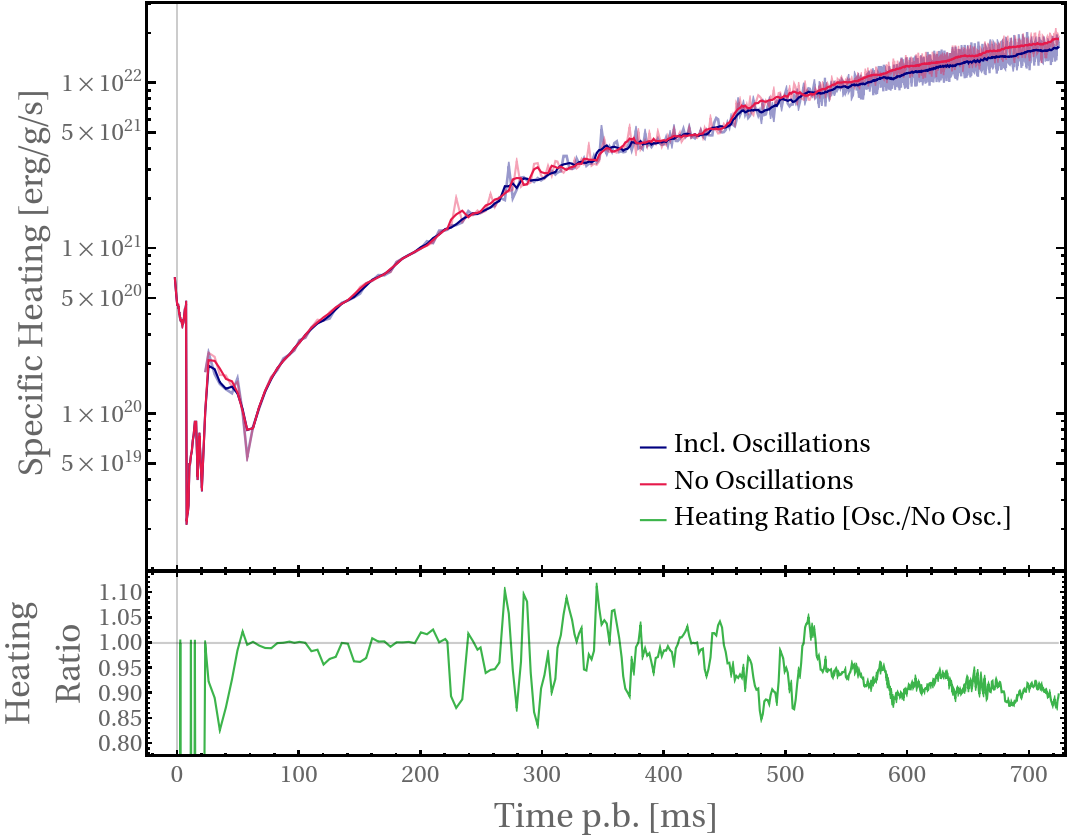}
    \caption{\label{fig:Heating} (Top) The net specific heating rate integrated over the gain region for both simulations. The bright lines show a time-averaged value of the instantaneous specific heating rate (faded lines) over 10ms. (Bottom) The ratio of average integrated specific heating with oscillations compared to the standard case with no oscillation.
    }
\end{figure}
These changes in the luminosities and mean energies will have a direct effect upon the net neutrino heating, an important quantity in the neutrino-driven supernova mechanism. In figure~\ref{fig:Heating} we show the net specific heating integrated over the gain region for both simulations. In the figure we see an initial small decrease in neutrino heating in the simulation with oscillations between $\sim20\,{\rm ms}$ and $\sim50\,{\rm ms}$. The heating rates then evolve similarly in both simulations for the next $\sim400\,{\rm ms}$, though with increasing variability starting between $\sim250\,{\rm ms}$ and $\sim300\,{\rm ms}$. Finally at $\sim 500\,{\rm ms}$ after bounce, there is another decrease in the heating rate due to the effect of neutrino oscillations.
\begin{figure}[h]
    \centering
    \includegraphics[width=0.45\textwidth]{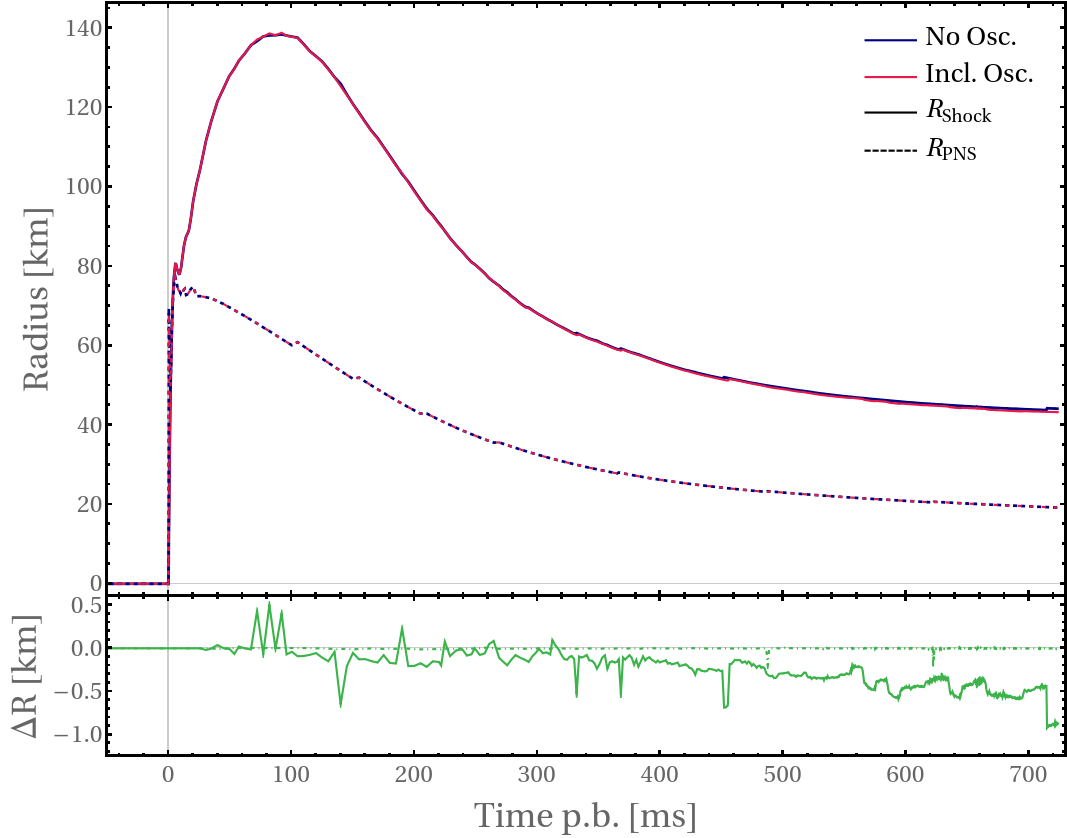}
    \caption{\label{fig:Rshock} (Top) The shock (solid) and proto-neutron star (dashed) radius for simulations with oscillations (orange) and with no oscillations (navy). (Bottom) The difference in PNS and shock radius between the two simulations, taken as $\Delta R = r_{\mathrm{osc.}}-r_{\mathrm{no\,osc.}}$
    }
\end{figure}

In figure~\ref{fig:Rshock}, the position of the shock and protoneutron star radius are shown, as well as the difference in their position. The overall behavior of both simulations is typical for non-exploding 1D models where the shock stalls at $\sim150$ km and then retreats before settling into a prolonged accretion phase. The shock position for both simulations is very close until $\sim 400$~ms after bounce. After this, the decrease in neutrino heating due to neutrino oscillations results in less support for the shock and a growing difference in shock position as shown in the bottom panel of figure~\ref{fig:Rshock}.
\newline


{\bf Conclusions} In this paper we have presented a new core-collapse supernova simulation code which includes hybrid classical-quantum neutrino transport. The coupling of the classical and quantum neutrino transport components was achieved using effective opacities which are updated as the simulation proceeds. We then presented results from a simulation using the new code and a 20 $M_{\odot}$ progenitor in order to explore the feedback of flavor transformation upon the explosion. We have found that the effects of the neutrino oscillations alter the neutrino luminosities and mean energies by a few percent which caused a decrease in neutrino heating below the shock. The change in neutrino heating was not enough to dramatically alter the outcome of the simulation of this particular progenitor. Our results agree with the previous studies about the potential for oscillations to alter the hydrodynamics \cite{2011PhRvL.107o1101C,2011ApJ...738..165S,2012PhRvD..85f5008D}, but with greater certitude. Whether this is also true for other progenitors is a topic for future investigation. 

Several improvements to the approach presented in this paper should be pursued. First, our treatment of the neutrinos as quantum particles applies only for the outgoing free-streaming neutrinos beyond the neutrinosphere. Close to the shock where we find the largest changes, this is a plausible approximation but clearly it is an approximation that needs to be relaxed in the future. Another major assumption is the use of the single-angle approximation in order to reduce the computational burden of the flavor transformation calculation. So-called multi-angle calculations \cite{Duan:2006jv} increase the computational expense considerably (a single invocation would be expected to take of order $10^3 - 10^4$ CPU hours \cite{2008CS&D....1a5007D}, compared to $\sim1$ CPU hour for our single-angle calculation) but nevertheless, would need to be performed eventually. Multi-angle calculations would also allow us to examine the feedback of fast-flavor oscillations should they occur. Finally, our simulations assumed spherical symmetry; however, multi-dimensional simulations are necessary if one wishes to examine whether additional non-linear effects found in 2D and 3D simulations might amplify the effect of oscillations. It will be interesting to see how all these future improvements might alter our results. \\


{\bf Acknowledgements} We wish to thank Bronson Messer, Sherwood Richers, and Matthias Liebend\"{o}rfer for many useful discussions in the development of this project. This research was funded by the US Department of Education Graduate Assistance in Areas of National Need (GAANN) grant number P200A150035 and the US Department of Energy, Office of Science, Office of Nuclear Physics under Award DE-FG02-02ER41216.

\bibliography{references}

\end{document}